\documentclass{article}
\usepackage{PRIMEarxiv}
\usepackage{hyperref}       
\usepackage[utf8]{inputenc} 
\usepackage[T1]{fontenc}    
\usepackage{booktabs}       
\usepackage{nicefrac}       
\usepackage{fancyhdr}       
\usepackage{listings}

    \title{NLP Cluster Analysis of Common Core State Standards and NAEP Item Specifications 
    \thanks {We used edited versions of the Common Core State Standards and NAEP Item Specifications. The edited versions, which appear in the Appendices of the previous paper, should not be used to represent the corresponding verbatim material.\cite{CamilliNLP} Thanks to Larry Jones of Colorado Mesa University for comments on on our example involving the idea of hardness. } }

\author{
  Gregory Camilli \\
  Rutgers University New Brunswick \\
   \texttt{greg.camilli@gse.rutgers.edu} 
   \and
  \textbf{Larry E. Suter} \\
  University of Michigan \\
   \texttt{Larrysuter@me.com} \\
}

\begin{document}
\large
\maketitle

\begin{abstract} Camilli (2024) proposed a methodology using natural language processing (NLP) to map the relationship of a set of content standards to item specifications. This study provided evidence that NLP can be used to improve the mapping process. As part of this investigation, the nominal classifications of standards and items specifications were used to examine construct equivalence. In the current paper, we determine the strength of empirical support for the semantic distinctiveness of these classifications, which are known as "domains" for Common Core standards, and "strands" for National Assessment of Educational Progress (NAEP) item specifications. This is accomplished by separate k-means clustering for standards and specifications of their corresponding embedding vectors. We then briefly illustrate an application of these findings.
\end{abstract}

\section {Introduction}
In an earlier paper, a methodology using natural language processing (NLP) was proposed for mapping the relationship of a set of content standards to a set of item specifications.\cite{CamilliNLP} In that investigation, the construct equivalence of educational standards and item specifications was examined. In the current study, the focus is on the nominal classifications per se, which are known as "domains" for Common Core Standards, and "strands" for the National Assessment of Educational Progress (NAEP) item specifications.\cite{NAGB2021, NGA} Because the methodology is the same for examining standards and specifications, we refer to both simply as "elements" when no distinction is required. The current study aims to determine the degree to which the nominal classifications of elements coincide with their corresponding empirical structure. In particular, we examine whether these nominal classifications can be reproduced by empirical classifications based on semantic textual similarity. 

The remainder of this paper has the following structure. First, a brief introduction to NLP semantic similarity is given followed by a description of k-means clustering, which is the method used in this study to obtain empirical classifications. The relationship of nominal to empirical (i.e., clusters) groupings is then obtained. Finally, we examine the implications of select classification mismatches and provide an illustration of applying the lessons learned. Hopefully, this paper will be useful for future studies of standards and item specification studies.

\section{Background}
Educational content standards constitute the centerpiece of the American system of public education under federal and state policy. The rationale is that both tests and instruction should be aligned to standards for maximum effectiveness.\cite{Forte,Smith} However, while standards provide a blueprint of prioritized content, they are too broad for test development. For this reason, item specifications are developed to translate the standards into concrete recipes for producing test items and assembling them into assessments. Both standards and specifications are subdivided to ensure adequate conceptual coverage of multidimensional constructs like mathematics proficiency. As we shall see, such multilevel structures may significantly impact test development and score interpretation.

Much effort is expended to align item specifications to content standards, the "official" subdivisions or categories are seldom scrutinized empirically.\cite{Camilli,Camilli2019} These subdivisions may be presumed or created after the fact as a convenience. Traditionally, academic discussions or arguments about content standards have occurred primarily among diverse content experts, who sometimes work from different ideological perspectives. \cite{Alan} However, the data in question are text elements, and analysis based on NLP methods may contribute new possibilities for improving how standards and item specifications are structured.     

 Test-item dimensionality can be considered either from a constructivist or an empirical perspective. According to the former approach, content groupings are created a priori in the minds of mathematicians and mathematics educators. For these categories to be consistent, the elements within a category should be more similar to each other than those within a different category. This is similar to the idea of convergent and discriminant validity. \cite{CampbellFiske} The present study examines the consistency of categories by first obtaining an empirical grouping of elements based on embedding vectors for the elements and then comparing the nominal and empirical classifications. This approach focuses on whether the nominal categories are consistent, and whether empirical elements that fall outside their nominal category can be justified or explained. Whether the categories (nominal or empirical) are inherently correct or beneficial is a topic beyond the scope of this investigation.

\subsection{NLP Studies}
Several studies have used NLP methods to examine alignment issues, which is more generally referred to as \textit{content mapping}. \cite{Khan, zhou, butterfuss}. Camilli summarized some of this research in a previous paper that examined how NLP methods can be used to evaluate content mapping studies. \cite{CamilliNLP} The data sets used in the current study were also described, and a number of key issues in content mapping were identified. In particular, important modifications to the Common Core standards and NAEP item specifications were explained. \cite{NAGB2021, NGA}

 \begin{table}
\begin{center}
 \caption {Abbreviations }  
\begin{tabular}{@{}lll@{}}
\toprule

CCSS    &  CCSS-opal  & Operations and Algebraic Thinking              \\
Domains &  CCSS-numb  & Number and Operations in Base 10               \\
        &  CCSS-frac &  Number and Operations—Fractions               \\
        &  CCSS-meas  & Measurement and Data                           \\
        &  CCSS-geom  & Geometry                                       \\ \midrule
NAEP    &  NAEP-numb  & Number Sense, Properties, and Operations            \\
Strands &  NAEP-meas  & Measurement                                         \\
        &  NAEP-geom  & Geometry                                            \\
        &  NAEP-data  & Data Analysis, Statistics, and Probability          \\
        &  NAEP-alge  & Algebra                                             \\ 
\bottomrule
\end{tabular}
\end{center}
\end{table}

\section{Notes on NLP Embedding}
Cohesive chunks of text in the NLP literature are alternatively referred to as sentences, extended sentences, text, or statements.  The resulting sentence data are then vectorized to obtain numerical representations called embedding vectors (EV). The word embedding is used to suggest "meaning in context." Each sentence (or segment of text) can be represented with an EV of dimension length \textit{n} of semantic attributes. Subsequently, the EVs of each word in a sentence are combined into a single vector to represent the sentence (or larger segment of text) as a whole. Further statistical analysis was conducted on these to understand the substantive properties of text. More details on the preparation of text for analysis were given in the previous paper. \cite{CamilliNLP} 

\section{Methods}
The method used to study the structure of EVs in this study is k-means cluster analysis, making only the assumption that 5 clusters exist. Below, we refer to the clusters as empirical groupings in contrast to nominal groupings. Once obtained, elements of the nominal and empirical groupings can be cross-classified to examine consistency. Suppose the nominal classes can be wholly retrieved empirically. In that case, the semantic distinctiveness of the nominal structure is supported. With less than perfect consistency, examining  mismatches in the cross-classification may help to sharpen conceptual boundaries for further development.

\subsection{Data Preparation}
The Common Core standards are nested within 5 content domains containing the 34 standards: (Operations and Algebraic Thinking; Number and Operations in Base Ten; Number and Operations—Fractions; Measurement and Data; Geometry). The 49 NAEP item specifications are nested within 5 content domains (Number sense, properties, and operations; Algebra and Functions; Measurement; Data Analysis, Statistics, and Probability; Geometry). This layout is shown in Table 1 for easy reference.

For the Common Core, the transformed data consisted of a $34 \times 3000$ matrix in which the rows correspond to the EV dimension for the standards. For cluster analysis, the first task is to reduce redundancy by extracting the first $5$ principal components (PC) of the data matrix. The main idea of principal component analysis is to extract a small number of orthogonal variables that account for the lion's share of the variance of a set of original variables. Eventually, $4$ of these PCs were used, resulting in a $34 \times 4$ data set for analysis (one PC showed essentially no variation between domains). A similar procedure was used for the NAEP specifications, which resulted in a $49 \times 4$ data set. Note that the maximum number of PCs available was determined by the rank of the covariance matrix, in this case $34$ (for standards) or $49$ (for specifications). Consequently, the data reduction step appeared to be highly successful.

\subsection{K-Means Cluster Analysis}
K-means is a method of cluster analysis that proceeds in two steps. First, a number of proposed clusters is selected or randomly generated. Second, a set of seed centroids is provided for a set of observations (in this case, the PCs for standards or for specification elements). Third, each observation is assigned to the nearest cluster (i.e., centroid) based on Euclidean distance and cluster centroids are recalculated. This process is iterated to convergence during which elements may change cluster membership. Convergence is attained when centroids change negligibly, at which point within-cluster variances are minimized, or alternatively, between-cluster variances are maximized (because the total variance is fixed). 

The drawbacks to k-means clustering include the requirement of selecting a working number of clusters, a bias toward spherical within-cluster distributions, and the assumption that cluster do not overlap. However, k-means is conceptually simple and frequently used as a baseline to evaluate other methods of clustering for particular applications. We make no claim this method is optimal. However, the clustering results are consistent with the nominal structures, and the few mismatches that occur (between nominal and empirical groupings) have compelling substantive explanations. 

\section{Results}

\subsection{Descriptive Statistics}
The PC cluster means for both the CCSS and NAEP data matrices are given in Table 2.  Because the PCs were not rotated, their intercorrelations are $r = 0$, and  standard deviations are $\sigma = 1$. While the original EVs are uninterpretable, the PCs may have some substantive interpretations based on their patterns of means. For example, PC4 for the CCSS seems to be discriminating between Measurement and Data, on the one hand, and Geometry, on the other. Note the cluster order is the same as the nominal order in Table 1.

\begin{table}
\begin{center}
 \caption {PC Cluster Means }  
\begin{tabular}{@{}rcrrrr@{}}
\toprule
        &Cluster & PC1   & PC2 & PC3  &PC4   \\ \midrule
CCSS    & 1&   0.00 & 0.08 & 0.24  &-0.11  \\
        & 2&  -0.09 & 0.34 &-0.10  & 0.03  \\
        & 3&  -0.23 &-0.22 &-0.06  & 0.04  \\
        & 4&  0.49 &-0.14 &-0.17  &-0.28  \\
        & 5&  0.50 &-0.03 & 0.17  & 0.39  \\ \midrule
NAEP    & 1& -0.26  & 0.01 & -0.10 & 0.06  \\ 
        & 2& 0.21   &-0.44 & -0.08 & -0.10 \\
        & 3& 0.30   & 0.22 & -0.11 & 0.00  \\
        & 4&0.08   &-0.06 & 0.37  & 0.30  \\
        & 5&-0.13  & 0.08 & 0.25  &-0.25  \\ \bottomrule
\end{tabular}
\end{center}
\end{table}

For both CCSS and NAEP, 4 PCs were sufficient to provide the maximum accuracy for the cross-classification matrix. Additional PCs were examined, but these showed negligible between-cluster variation and did not improve classification accuracy. 

\subsection{ Cross-Classification: Common Core Standards}
In Table 2, the classification results are given for the similarity of CCSS domains and empirical clusters. For each domain, a clear matching cluster exists. There are 6 mismatches, resulting in a classification accuracy of $82.5 \%$. Three of the mismatches consist (in the first column of Table 3) of misclassifying the nominal standard as belonging to an empirical operations and algebra cluster. Thus, one fractions element and two measurement elements appear more like operations and algebra. 

\begin{table}
\begin{center}
   \caption {CCSS Classification Results}
\begin{tabular}{@{}llrrrrr@{}}
\toprule
\multicolumn{2}{l}{CCSS} &  \multicolumn{5}{c}{Cluster} \\  \cline{3-7}

Domain    &  & 1   & 2   & 3    & 4     &  5  \\ \midrule
CCSS-opal     &  & 4   & 1   & 0    & 0     &  0     \\
CCSS-numb     &  & 0   & 6   & 0    & 0     &  0      \\
CCSS-frac     &  & 1   & 1   & 11   & 0     &  0     \\
CCSS-meas     &  & 2   & 0   & 1    & 4     &  0    \\
CCSS-geom     &  & 0   & 0   & 0    & 0     &  3    \\ \bottomrule
\end{tabular}
\end{center}
\end{table}

\subsection{Cross-Classification: NAEP Item Specifications}
For each NAEP strand, a clear matching empirical cluster was also obtained. In Table 5, there are 4 mismatches, resulting in a classification accuracy of $91.8 \%$. Two measurement specifications were misclassified as belonging to the geometry strand.

\begin{table}
\begin{center}
   \caption {NAEP Classification Results}
\begin{tabular}{@{}llrrrrr@{}}
\toprule
\multicolumn{2}{l}{NAEP} &  \multicolumn{5}{c}{Cluster} \\  \cline{3-7}

Strand    &  & 1   & 2   & 3    & 4     &  5  \\ \midrule
NAEP-numb     &  & 18  & 0   & 0    & 0     &  0     \\
NAEP-meas     &  & 0   & 7   & 2    & 0     &  0      \\
NAEP-geom     &  & 0   & 0   & 9    & 0     &  0     \\
NAEP-data     &  & 0   & 0   & 0    & 4     &  0    \\
NAEP-alge     &  & 0   & 0   & 1    & 1     &  7    \\ \bottomrule
\end{tabular}
\end{center}
\end{table}

\section{Classification Errors}
A brief review is given below for the cross-classification mismatches. The goal is to determine whether the commonalities among the misclassifications make sense. Because NLP methods may provide misleading results, misclassifications require  substantive examination. This review process may result in defensible explanations, leading to better understanding of how standards or item specifications are structured. 

\subsection{CCSS}

 Select mismatches from the CCSS cross-classification are shown in the top section of Table 5. In particular, we examine the $3$ Measurement and Data standards that are misclassified as operations and algebraic thinking.

\begin{quote}
\begin{description}
\item [4.MD.A.1] Know relative sizes of measurement units within one system of units including kilometers, meters, centimeters; kilograms, grams; pounds, ounces; liters, milliliters; hours, minutes, seconds. Within a single system of measurement, express measurements in a larger unit in terms of a smaller unit. Record measurement equivalents in a two-column table. For example, know that 1 foot = 12 x 1 inch. Express the length of a 4 foot snake as 48 inches. Generate a conversion table for feet and inches listing the number pairs (1, 12), (2, 24), (3, 36), (4,48).
\item [4.MD.A.2.] Use the four operations to solve word problems involving distances, intervals of time, liquid volumes, masses of objects, and money, including problems involving simple fractions or decimals, and problems that require expressing measurements given in a larger unit in terms of a smaller unit. Represent measurement quantities using diagrams such as number line diagrams that feature a measurement scale.
\item [4.MD.A.3.] Apply the area and perimeter formulas for rectangles in real world and mathematical problems. For example, find the width of a rectangular room given the area of the flooring and the length, by viewing the area formula as a multiplication equation with an unknown factor.
\end{description}
\end{quote}

These standards mainly involve operations and algebraic thinking, but they also involve specific content knowledge of how measurements relate to one another. The issue here is whether the latter consists merely of memorization facts (e.g., a kilo is equal to 2.20462 pounds) appended to number and operations skills. Linking measurement proficiency to formulaic conversions is a common practice in establishing content standards in mathematics. 

\subsection{NAEP}
The mismatches from the NAEP cross-classification are shown in the bottom section of Table 5. Here, we examine the 2 "Measuring Physical Attributes" standards that are misclassified as belonging to geometry.

\begin{quote}
\begin{description}
\item [4.Measuring Physical Attributes(f)] Solve problems  involving perimeter of plane figures.
\item [4.Measuring Physical Attributes(g)] Solve problems involving area of squares and rectangles. 
\end{description}
\end{quote}

These measurement standards clearly center on geometric concepts. Measurement is involved only in the sense of calculating perimeters or areas of geometric figures. We would argue that this is not measurement in the sense of the CCSS. The problem here is that the word "measurement" can have different meanings. The CCSS and NAEP appear to take different perspectives.

\begin{table}
\begin{center}
 \caption {Classification Errors }  
\begin{tabular}{@{}lll@{}}
\toprule
Common Core      &                                      &                                  \\
Standard         & Domain &  Mismatch                                                      \\ \midrule
        4.OA.B.4 &  Operations and Algebraic Thinking  & Number and Operations in Base 10  \\
        4.NF.C.7 &  Number and Operations—Fractions    & Number and Operations in Base 10  \\
        4.MD.A.1 &  Measurement and Data               & Operations and Algebraic Thinking \\
        4.MD.A.2 &  Measurement and Data               & Operations and Algebraic Thinking \\
        4.MD.A.3 &  Measurement and Data               & Operations and Algebraic Thinking \\  
        4.MD.B.4 &  Measurement and Data               & Number and Operations—Fractions   \\  \midrule
NAEP             &                                     &                                   \\
Specification           & Strand &  Mismatch                                               \\ \midrule
        4.Measuring Physical Attributes(f) &  Measurement   &  Geometry  \\
        4.Measuring Physical Attributes(g) &  Measurement   & Geometry \\
        4.Patterns, Relations, and Functions (a) &  Algebra & Geometry  \\
        4.Patterns, Relations, and Functions (d) &  Algebra & Data Analysis, Statistics \& Probability  
        \\  \bottomrule
\end{tabular}
\end{center}
\end{table}
\newpage
\section{Digression on Measurement}
The NLP cluster analysis has shown that semantically-speaking, there appears to be overlap between what is considered measurement, on the one hand, and either geometry or algebra, on the other. In fact, the word \textit{measurement} commonly refers to a wide variety of topics in the field of educational testing, ranging from counting to ordering to scale conversions to determining geometric quantities. This approach to measurement is described next, followed by brief consideration a more conceptual approach.

\subsection{Topic-based Measurement}
Especially in educational testing, measurement has been operationalized as a list of topics including: 

\begin{itemize}
\item physical attributes, like temperature, length, mass
\item geometric features, like length, height, width, area, circumference, volume
\item spatial relations, angles, graphs
\item comparison, ordering, transitivity (a < b and b < c implies a < c)
\item units of measurement, conversion of units within and across scales
\item data, estimation, precision, measurement instruments
\item counting, systems of whole numbers, benchmarks
\end{itemize}

This is only a small sample of potential topics. A more detailed overview of measurement is provided in the NAEP 2026 mathematics framework: 
\begin{quote}
   The connection between measuring and number makes measurement a vital part of school mathematics. Measurement is an important setting for negative and irrational numbers as well as positive numbers, since negative numbers arise naturally from situations with two directions and irrational numbers are commonplace in geometry. Measurement representations and tools are often used when students are learning about number properties and operations. For example, area grids and representations of volume using unit cubes can help students understand multiplication and its properties. The number line can help students understand ordering and rounding numbers. Measurement also has a strong connection to other areas of school mathematics and other subjects. Problems in algebra are often drawn from measurement situations and functions are used to relate measures to each other. Geometry regularly focuses on measurement aspects of geometric figures. Probability and statistics provide ways to measure chance and to compare sets of data. The measurement of time, values of goods and services, physical properties of objects, distances, and various kinds of rates exemplify the importance of measurement in everyday activities. (pp. 23-24) \cite{NAGB2021}
\end{quote}

Given this variety of topics, one line of reasoning might be that proficiency in measurement is simply shorthand for proficiency across a select set of numerical skills.  Consider finding the perimeter of a triangle, or converting inches to feet. These are commonly thought of in terms of measurement, yet there is no obvious connection except that both require a numerical result. Ultimately, this distinction may not matter. So-called measurement skills are always an aspect of one or more content domains. For example, measuring an angle is a skill relevant to geometry and converting units is relevant to mathematical transformations. If a test includes material on geometry and algebra, then no harm results from use of the term "measurement" as a label. However, when a distinct score is reported for measurement, it can only be interpreted relative to the particular topics included in the measurement domain. 

A second line of reasoning is that a list of topics, however, organized, is a conceptually bereft approach to measurement. Below, we attempt to provide a broad outline of measurement as a distinct discipline that provides conceptual tools to diverse fields such as (but not limited to) astronomy, forensics, education, medicine, archaeology, financial analysis. 

\subsection{Conceptual Measurement}
The 2026 framework of the National Assessment of Educational Progress (NAEP) provides a description of conceptual measurement that attempts to weave disparate topics together:

\begin{quote}
 Measuring is the process by which numbers are assigned to describe the world quantitatively. This process involves selecting the attribute of the object or event to be measured, comparing this attribute to a unit, and reporting the number of units” (p. 23).\cite{NAGB2021}   
\end{quote}

According to this definition, the construct of measurement involves mastery of a set of related principles, including the selection of an attribute, defining the unit of measurement, and assigning a certain number of units to an object based on the corresponding values of the attribute. There are other important features of measurement  (e.g., instruments, standardization, and evaluation of uncertainty), but the focus of measurement is on an attribute or on a construct. \textit{Attribute} is used to designate an aspect of a physical thing (e.g., temperature, length, mass), whereas an aspect of human beings (e.g., mathematics proficiency or extroversion) is typically designated as a \textit{construct}. Attributes or constructs are features of objects or people, respectively, that are of interest. A physical object can have more than one feature and the same is obviously true of human beings: a person may be both intelligent and  annoying.

\subsubsection{Illustration}
 An example outside the field of education may provide a fresh perspective on measurement as a conceptual process. Consider two scales used to describe the hardness of a physical material:

\begin{itemize}
    \item The Vickers procedure uses a diamond to indent the surface of a metal, and the force exerted is divided by the surface area of the indentation and transformed to scale values. The attribute being measured is resistance to indentation. With this scale you can assign quantitative values to metals, and define a unit of measurement that in turn can be used to compare materials (e.g., one metal may be twice as hard as another). This procedure is carried out with a precision measuring instrument.
    \item The Mohs procedure uses a manual scratch test. If mineral A scratches mineral B, and mineral B does not scratch A, then then A is harder than B. The test is then applied with reference to a select set of 10 benchmark minerals, which are subsequently ranked from 1 (softest, talc) to 10 (hardest, diamond). The attribute being measured is scratch resistance visible to the naked eye. The hardness of minerals can subsequently be determined in relation to these benchmarks. 
\end{itemize}

Both scales involve the idea of arranging materials with respect to hardness, but differ in two important ways. First, resistance to indentation for metals is not the same thing as scratch resistance for minerals. In fact, the term “hardness” is not a single entity. Vickers suffices for some applications (e.g., testing hardened steel in the lab), Mohs suffices for others (e.g., evaluating minerals in the field), and still other hardness scales exist for other purposes. In contrast to the Vickers, Mohs produces only an ordering based on relative hardness rather than a quantitative value, consequently there is no “unit” of measurement. For this reason, differences between or ratios of Mohs values are not meaningful. The Mohs scale could be based on letters, A, B, C ..., and the utility and meaning of the scale would not change. 

This example suggests there are fundamental processes are involved in making sense out of variability in hardness. First, an attribute must be chosen, but this choice depends on the application because attributes don’t exist without purpose (this doesn’t imply the object of measurement is arbitrary). In turn, the standardization of attributes, instruments, and procedures is required for determining reliability. Second, some scales have scalable “units” (Vickers) and some don’t (Mohs). There are scholars who believe that if there is no unit, there is no measurement. Taken literally, however, this would leave us without a good word for describing the Mohs procedure. Indeed, most uses of the word measurement would need to be stricken from the English language. In any case, the qualitative contrast between the Vickers and Mohs scales is useful because it helps us to think about how to specify (or theoretically verify) attributes or constructs, establish useful proxies for ideal measurement, and improve data quality. 

\subsubsection{Measuring Measurement}
Measuring proficiency in the construct of measurement would be challenging because there is no popular consensus on what measurement is, despite a cornucopia of available definitions. Moreover, it is not clear that there is a need for assessing this construct--with two caveats. First, the case can still be made that teaching measurement theory would be beneficial. Certainly measurement can be taught from a technical perspective at the post-secondary level (e.g., econometrics or psychometrics), but the core principles are basic to scientific method, including how attributes and constructs are identified, distinguished, and investigated. The example above hints that conceptual measurement would be most effectively taught in case studies or projects involving substantive applications. Second, the potential lack of cohesion in topic-based measurement may lead to fuzziness in both mathematics standards and item specifications. This may impart difficulties to both standards-based instruction and test development.

\section{Discussion}
In terms of semantic textual similarity, the Common Core standards and NAEP item specifications are internally consistent ($83\%$ and $92\%$, respectively) but a handful of mismatches unidentified raise questions about how the idea of measurement is implemented. As pointed out above, this issue involves the myriad connotations of the word measurement in the English language. Even experts in the field of psychometrics cannot agree upon a definition of measurement.  From a more pragmatic perspective, an effort to distinguish or classify topics  (e.g., measurement, geometry, and numeric operations) may reduce redundancies in test construction or overlap in test subscores. Whether and where in the curriculum students should be taught a conceptual understanding of measurement and its role in the sciences is an ongoing question.\cite{Maul} \cite{Newton,Borsboom,Wilson}

Finally, it is important to recognize that traditional document analysis would have plausibly revealed the same findings as reported here. Still, an NLP application  produces faster results with fewer resources. These results, in turn, be used to make subject matter deliberation more efficient and cost effective. In any case, the issues raised in this paper are emblematic of how NLP can be used to gain new insights into established measurement practices. 

\newpage

\bibliographystyle{unsrt}  
\bibliography{main}  

\end{document}